\documentclass[letterpaper, 10 pt, conference]{ieeeconf}  % Comment this line out if you need a4paper
\usepackage{graphicx}
\usepackage{tabularx}
\usepackage{multirow}
\usepackage[normalem]{ulem}
\useunder{\uline}{\ul}{}
\newcolumntype{L}[1]{>{\raggedright\let\newline\\\arraybackslash\hspace{0pt}}m{#1}}
\newcolumntype{C}[1]{>{\centering\let\newline\\\arraybackslash\hspace{0pt}}m{#1}}
\newcolumntype{R}[1]{>{\raggedleft\let\newline\\\arraybackslash\hspace{0pt}}m{#1}}
\newcolumntype{Y}{>{\centering\arraybackslash}X}

\IEEEoverridecommandlockouts                              % This command is only needed if 
\title{\LARGE \bf
AiAReSeg: Catheter Detection and Segmentation in Interventional Ultrasound using Transformers}

\author{Alex Ranne$^{1}$, Yordanka Velikova$^{2}$, Nassir Navab$^{2}$, and Ferdinando Rodriguez y Baena$^{1}$% <-this % stops a space
\thanks{This work was supported by the UKRI CDT in AI for Healthcare under Grant EP/S023283/1, the ICL-TUM Joint Academy of Doctoral Studies (JADS) program, and the TUM Global Incentive Fund.
}% <-this % stops a space
\thanks{$^{1}$Alex Ranne and Ferdinando Rodriguez y Baena are with the Hamlyn Centre for Robotic Surgery, Imperial College London, SW7 2AZ, UK   (e-mail: {\tt\footnotesize
 \{alex.ranne17, f.rodriguez\}@imperial.ac.uk}) (Corresponding author: Alex Ranne).}
\thanks{$^{2}$Yordanka Velikova and Nassir Navab are with the Chair for Computer Aided Medical Procedures
and Augmented Reality (CAMP), Technical University of Munich, 85748
Garching, Germany (e-mail: {\tt\footnotesize
 \{dani.velikova, nassir.navab\}@tum.de}) %
}}

\begin{document}

\markboth{Robotics \& Automation Letters, March~2023}%
{Ranne \MakeLowercase{\textit{et al.}}: AiAReSeg: Catheter Detection and Segmentation in Interventional Ultrasound sequences using Transformers}

\maketitle
\thispagestyle{empty}
\pagestyle{empty}

% \setlength{\abovedisplayskip}{0pt}

%%%%%%%%%%%%%%%%%%%%%%%%%%%%%%%%%%%%%%%%%%%%%%%%%%%%%%%%%%%%%%%%%%%%%%%%%%%%%%%%
\begin{abstract}

To date, endovascular surgeries are performed using the golden standard of Fluoroscopy, which uses ionising radiation to visualise catheters and vasculature. Prolonged Fluoroscopic exposure is harmful for the patient and the clinician, and may lead to severe post-operative sequlae such as the development of cancer. Meanwhile, the use of interventional Ultrasound has gained popularity, due to its well-known benefits of small spatial footprint, fast data acquisition, and higher tissue contrast images. However, ultrasound images are hard to interpret, and it is difficult to localise vessels, catheters, and guidewires within them. This work proposes a solution using an adaptation of a state-of-the-art machine learning transformer architecture to detect and segment catheters in axial interventional Ultrasound image sequences. The network architecture was inspired by the Attention in Attention mechanism, temporal tracking networks, and introduced a novel 3D segmentation head that performs 3D deconvolution across time. In order to facilitate training of such deep learning networks, we introduce a new data synthesis pipeline that used physics-based catheter insertion simulations, along with a convolutional ray-casting ultrasound simulator to produce synthetic ultrasound images of endovascular interventions. The proposed method is validated on a hold-out validation dataset, thus demonstrated robustness to ultrasound noise and a wide range of scanning angles. It was also tested on data collected from silicon-based aorta phantoms, thus demonstrated its potential for translation from sim-to-real. This work represents a significant step towards safer and more efficient endovascular surgery using interventional ultrasound.

\end{abstract}

%%%%%%%%%%%%%%%%%%%%%%%%%%%%%%%%%%%%%%%%%%%%%%%%%%%%%%%%%%%%%%%%%%%%%%%%%%%%%%%%
\section{INTRODUCTION}

Cardiovascular disease is the most common cause of death in the world, accounting for 17.9 million deaths per annum \cite{kaplan2017kaplans}. Traditionally, open surgery is performed to expose the diseased vasculature, which poses significant trauma for the patient. As an alternative, computer-assisted minimally invasive endovascular surgery has been widely adopted due to its benefits of reducing patient recovery time, and lower risk of infection, thus saving costs for healthcare providers, and more importantly saving lives.

\begin{figure}[t!]
    \centering
    \includegraphics[width=\columnwidth]{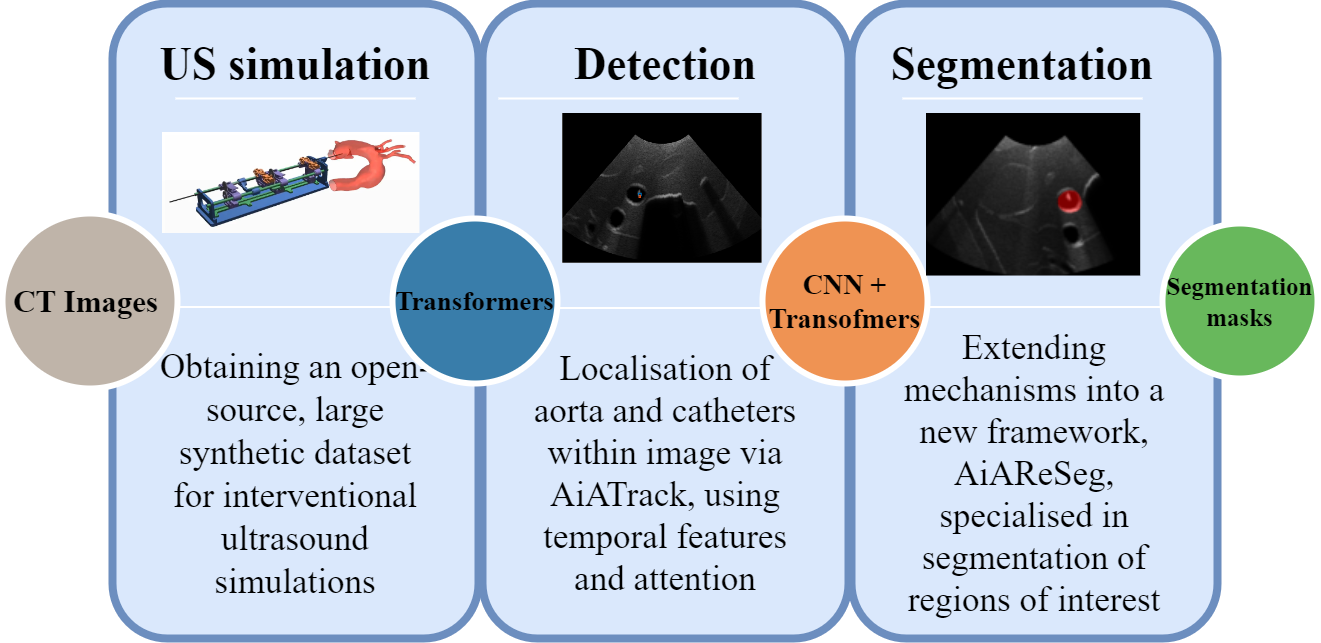}
    \caption{Main workflow pipeline of proposed system. Stage 1: Data synthesis via physics engine and ray casting. Stage 2: Detection of critical anatomy locations. Stage 3: Semantic segmentation}
    \vspace{-8mm}
    \label{Fig:Main}
\end{figure}

In endovascular surgery, catheters and guidewires are steered, under Fluoroscopic guidance, through torturous vessel trees to reach their desired destination \cite{abdelaziz2021x}. During navigation, staff and patient are exposed to prolonged periods of ionising radiation, which increases the risk of developing cancer. During an intervention, in order to visualise the vessels, the patient is also injected with a radiopaque dye (Digital Subtractive Angiography, DSA), which is harmful for the kidneys. On the other hand, this system still lacks the ability to obtain feedback on real-time instrument positions relative to the vasculature. This may introduce additional risks for the patient, as there may be frequent and unintentional contacts between these instruments and the vessel wall, with the consequent risks of perforation, dissection, thrombosis and embolization \cite{walker2010clinical}.

Alternatively, intraoperative Ultrasound imaging (iUS) offers a non-ionising solution for visualisation. In comparison to Fluoroscopy, US imaging has been a popular tool in diagnosis and aneurysm screening \cite{ullery2018epidemiology}, due to the high tissue contrast, temporal resolution, and efficacy \cite{szabo2004diagnostic}. In surgery, clinicians have applied it in conjunction with or completely replacing Fluoroscopy, in endovascular aneurysm repair \cite{von2002routine, kopp2010first}, Balloon Angioplasty \cite{wakabayashi2013ultrasound}, and Electrophysiology \cite{mangrum2002intracardiac}.

In order to monitor the instruments, the surgeon must detect the tip position of the catheter in US images, which poses a significant challenge for them. To begin with, the spatial resolution of an ultrasound image is limited by the number of elements in the transducer, and by the trade-off relationship with the penetration depth \cite{hartshorne2011ultrasound}. In order to examine deep into the target tissue, the clinician must lower the frequency as waves with higher wavelength experience less attenuation \cite{gibbs2011ultrasound}. However, this is at the expense of spatial resolution. Secondly, the noisy nature of ultrasound makes interpretation of images difficult, since images contain clutter, shadowing, and reverberation artifacts. Consequently, labeling and interpreting the image requires expert knowledge in order to relate the physical anatomy with the image, which may vary in quality, resolution, intensity, and acquisition protocols, and are not standardized. The progress in deep learning for US instrument detection and segmentation in endovascular procedures offers an opportunity for the field to reform. With the development of object detection networks, researchers have identified their potential in finding objects of varying sizes, and streamlining their workflow. Designing a suitable architecture for this task, and acquiring sufficient data to do so are two challenges that need to be solved.

In this paper, we propose a novel three-step framework to overcome the lack of intraoperative ultrasound data of a cathetarisation, required to train our network. In Sect. \ref{Methodology}, we propose to generate synthetic iUS data with instruments inside by fusing a physics engine into an existing CT-to-US simulator, thus generating mechanically realistic scans. Once generated, this data was then used to train a novel detection and segmentation architecture (AiAReSeg), which we propose in in Sect. \ref{Methodology: AiAReSeg Architecture}. Finally, in Sect. \ref{Experimental evaluation}, we evaluate the trained model on a hold out validation set of simulated US data, but also with aortic phantom images, which resemble a true surgical environment to a closer degree. 

\section{Related works}

\subsection{State of the art deep learning architectures}

\subsubsection{Detection}
In terms of architecture, most popular networks fall into two categories: convolution-based (CNN) or attention-based. In CNN-based systems, Faster-R-CNN \cite{ren2015faster} leveraged the power of its region proposal network (RPN) to select regions of interest, prior to passing such features into fully connected layers for bounding box prediction. The network achieved real-time performance since the need for hand-picked anchor points (found in Fast-R-CNN \cite{girshick2015fast}) was removed. However, following the introduction of attention in the Transformer architecture \cite{vaswani2017attention}, vision transformers became a strong contender for CNNs as they can learn global dependencies from across the image with a patch-based approach, then concatenating the attention maps together to form the final prediction \cite{dosovitskiy2020image}. Much more recently, researchers have continued to evolve the field by combining the benefits of both worlds, fusing a CNN feature extractor with the attention mechanism. From this idea emerged numerous variants of the transformer, such as the Detection Transformer (DETR) \cite{carion2020end}, which used a ResNet50 backbone for feature extraction, before feeding its output into a transformer that provided embeddings corresponding to various objects in the scene. Using the bipartite matching loss \cite{stewart2016end}, the network minimised the difference between a prediction output and the ground truth in a class-specific manner.

\subsubsection{Semantic Segmentation}
In the context of semantic segmentation, the CNN-based UNet \cite{ronneberger2015u} architecture and its adaptations have also performed exceptionally well in segmentation tasks. Following the introduction of the nnUNet pipeline \cite{isensee2021nnu}, which proposed an all-in-one pre-processor, parameter and model selection pipeline, the performance of UNet has been further refined. With that said, nnUNet does not operate in real-time, making it not suitable for high-speed US. On the other hand, attention-based segmentation networks have also seen much success, such as with the DETR to perform panoptic segmentation tasks \cite{carion2020end}, or in the case of the Segmentation Transformer (SETR) \cite{zheng2021rethinking}, which removed the ResNet50 backbone, but used a Sigmoid activation function to generate segmentation masks.

\subsubsection{AiATrack - Learning with temporal features}

Thus far, aforementioend models only use spatial features learned in a single frame to make the prediction. This may be sufficient for good-quality images, but may fail when there is occlusion due to shadows or artifacts. To solve this problem, we drew inspiration from clinicians, who rely on prior knowledge from the previous position of the aorta to reposition the probe and relocate the lost targets. This concept was previously captured in the AiATrack framework \cite{gao2022aiatrack}, where a ResNet50-Transformer framework was used together with a corner-predictor based bounding box head. However, the final box prediction still only draws information from the transformer decoder outputs, instead of across the entire sequence of data. We believe we can further improve this invention to operate on even more challenging tasks, such as locating a small catheter's cross-section in sequences of highly variable US images.

\subsection{State of the art in US image simulation}

\begin{figure*}[t!]
    \centering
    \scalebox{0.75}{
    \includegraphics[width=\textwidth]{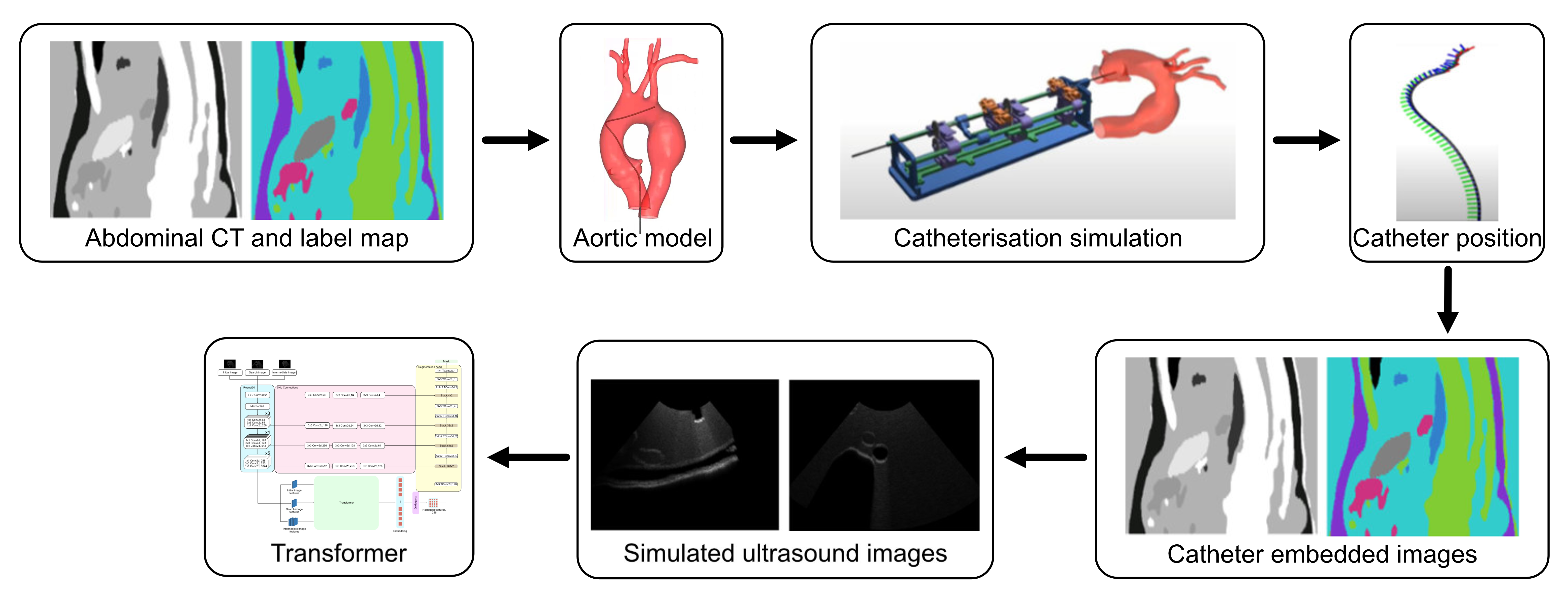}}
    \vspace{-5mm}
    \caption{The pipeline for generating synthetic ultrasound images from open source CT volumes. The aorta model is initially extracted, imported into the catheterisation engine, with the catheter positions exported as a time series, then redrawn on the CT images, before passed into a ray-tracing based simulator to generate the finished image}
    \vspace{-7mm}
    \label{Fig:CathSim_pipeline}
\end{figure*}

Despite the impressive results achieved by deep learning, the majority of network architectures are supervised, learning from an extensive number of labeled ground truths, which for Ultrasound is not readily available due to the difficulties faced in acquisition. However, there are publicly available large sets of pixel-level labeled CT volumes, which can be translated into simulated US image/label pairs \cite{rubi2017comparison} where further data augmentation can be added via applying rotation, brightness jitter, random shadowing, and artificial tissue deformations, etc. In this way, a large training dataset can be generated for the pretraining of deep learning architectures, allowing the networks to learn domain-specific feature extraction, before retraining on a significantly smaller, real dataset. Evidence of successful transfer shown previously in Velikova et al.'s work \cite{velikova2022cactuss, salehi2015patient} motivates this idea.

\section{Methodology} \label{Methodology}

\subsection{CT Data selection} \label{Methodology: Data selection}
In this work, labeled CT volumes of 8 men and women were acquired from the publicly available dataset Synapse \footnote{https://www.synapse.org//\#!Synapse:syn3193805/wiki/89480}. The labels of bone, fat, skin and lungs were added in the label map to allow for the simulation to function. The detailed process of generating the interventional US is detailed below.

\subsection{Physics-based catheterisation simulation} \label{Methodology: US Simulation}

Since obtaining a large dataset for the initial training of a deep learning architecture is time consuming, we are proposing a new US data simulation pipeline for generating interventional data, which is otherwise only attainable from the operating room environment or via a phantom. 

In literature, there are two ways to generate US data: finite difference solutions of the wave equation \cite{jensen1997new, treeby2020nonlinear}, or ray-casting through an image volume, semantically labelled with its acoustic attenuation properties \cite{salehi2015patient}. Since the former method requires solving a large system for each frame, it typically requires significantly longer computation time. Thus, the second method, albeit not as accurate, was selected. The simulator selected uses a hybrid ray-tracing convolutional method to define an anatomical representation that mimics the texture of real US images, define anisotropic properties, generate artifacts, and provide tissue contrast that allows regions of interest to be easily discernable. 

In order to generate a dataset consisting of catheters, we repurposed an open source catheterisation simulator developed by Jianu et al. \cite{jianu2023cathsim}, which is able to recreate high fidelity catheter-aorta mechanical interaction simulations. CathSim is built in the MuJoCo physics engine (DeepMind, London, UK) \cite{todorov2012mujoco, jianu2023cathsim}, which is a powerful package that can perform real time multi-joint dynamics computations with contacts using a C based API.

The final preprocessing pipeline is detailed in Fig. \ref{Fig:CathSim_pipeline}. CathSim renders the mesh models of the aorta and the catheter separately, while the aorta mesh was divided into 1024 convex hulls, decomposed using the V-HACD algorithm. The decomposed hulls were transformed into the same coordinate frames as the simulated environment via Blender v3.2.1 (Blender Foundation, Amsterdam, Netherlands) and imported into CathSim. The insertion simulation was performed with a linear translation speed of 0.1m/s, and inserted for 1,000 time increments, where each increment represents 1/60th of a second, and positions of the catheter were sampled at regular intervals along its body, and exported into a time series csv, which was transformed back into the CT's coordinate system. Finally, the simulator was initialised with multiple splines on the surface of the patient's torso, and the splines were tilted at angles of 0, $\pm$ 30 and $\pm$ 60 degrees, and a sweep of 1,000 images were generated for each angle.
\vspace{-5mm}
\subsection{AiAReSeg Architecture}\label{Methodology: AiAReSeg Architecture}

\begin{figure}[b!]
    \centering
    % \vspace{-5mm}
    \scalebox{0.95}{
    \includegraphics[width=\columnwidth]{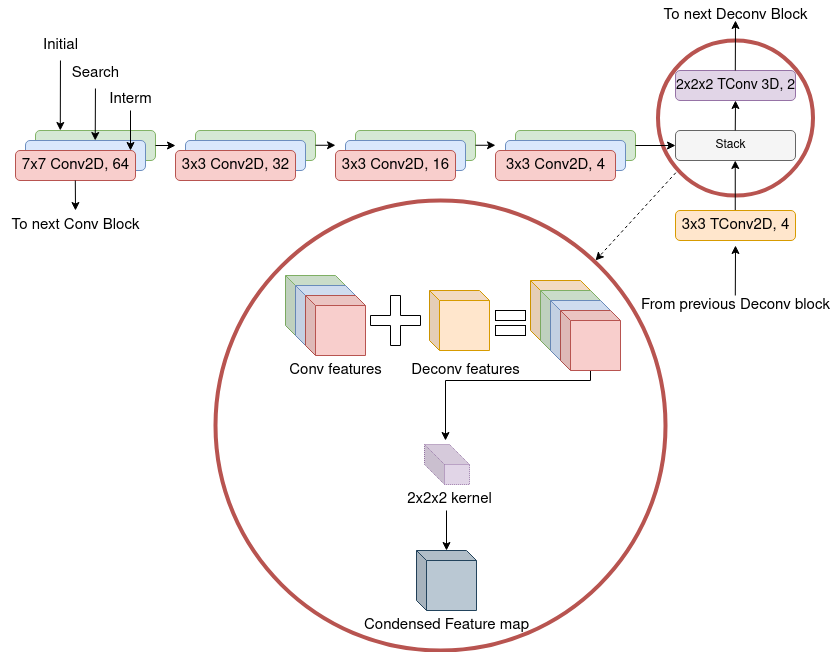}}
    \caption{Closeup of the 3D deconvolution pipeline. The three input feature maps are color coded in green, blue, and green for the initial, search and intermediate frames, respectively. The three frames are stacked with the output from the previous deconvolutonal block (amber), then deconvolution is performed with a 3D 2x2x2 kernel.}
    \label{fig:3D deconv}
\end{figure}

% \vspace{-5mm}
\begin{figure*}[t!]
    \centering
    \scalebox{0.8}{
    \includegraphics[width=\textwidth]{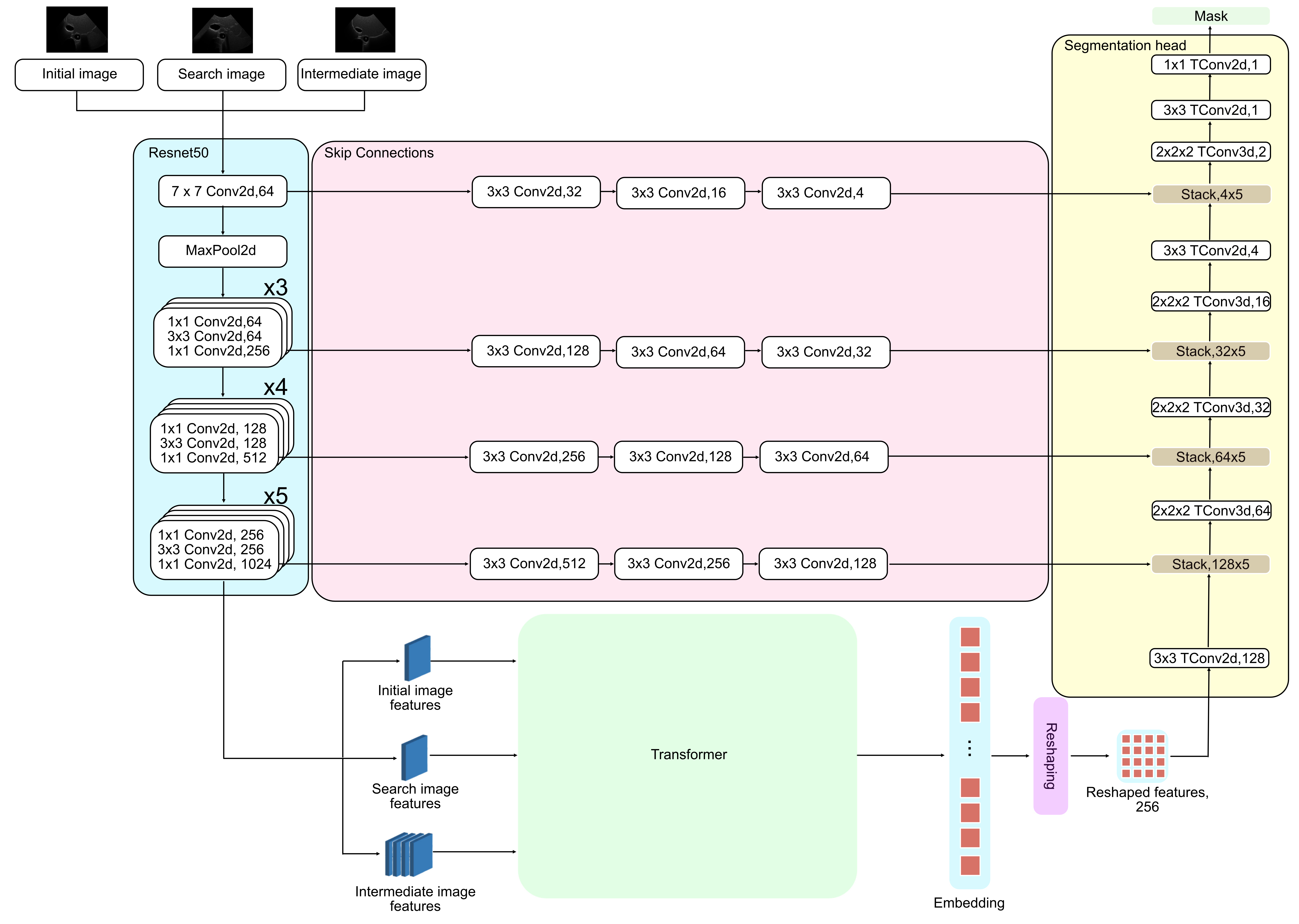}}
    \vspace{-5mm}
    \caption{The AiAReSeg architecture. Details of each module and its channel number is shown in the figure.}
    \vspace{-5mm}
    \label{fig:AiAReSeg}
\end{figure*}

Attention in Attention + ResNet for Segmentation (AiAReSeg) is a novel segmentation architecture that is adapted from AiATrack \cite{gao2022aiatrack}, shown in Fig.\ref{fig:AiAReSeg}. The main architecture consists of three components, the attention-in-attention module, the transformer architecture, and the outer upconvolution-deconvolution layers. 

Attention-in-attention (AiA) was first proposed by Gao et al.'s work \cite{gao2022aiatrack}, where the authors observed that each query-key pair generated an independent attention map, which ignored the features of other maps. The original attention mechanism used the following dot-product equation:
\vspace{-1mm}
\begin{equation}
    Attn(\mathbf{Q},\mathbf{K},\mathbf{V}) = (\textup{Softmax}\left ( \frac{\bar{\mathbf{Q}}\bar{\mathbf{K}}^{T}}{\sqrt{C}} \right )\bar{\mathbf{V}})\mathbf{W_o}
\end{equation}

Where $\bar{Q} = QW_q$, $\bar{K} = KW_k$, $\bar{V} = VW_v$, where $Q,K,V$ are the queries, keys and values, respectively, while $W_q, W_k, W_v$ denotes the learnable weight arrays for the query, keys and values, and $C$ is the channel size. In the case of a noisy dataset with distracting backgrounds, the model may become confused from clutter in the scene, leading to poor predictions. However, it was noticed that the attention weights near regions of interest were significantly higher, and pixels in such regions were of more interest than pixels with high attention weights further away. Thus, the designed AiA module applied attention once again on the attention map $M$ to filter out distant weights, which can be represented as:

\vspace{-5mm}
\begin{equation}
    InnerAttn(\mathbf{M}) = (\textup{Softmax}\left ( \frac{\bar{\mathbf{Q'}}\bar{\mathbf{K'}}^{T}}{\sqrt{D}} \right )\bar{\mathbf{V'}})\mathbf{(1+W'_o)}
\end{equation}
Where $\bar{Q'}, \bar{K'}, \bar{V'}$ are intermediate weighted queries, keys and values, which were feature vectors taken from columns in $M$, while $D$ represents the intermediate channel size, defined in this case as the height of the attention map.

When combined, the AiA module computes the following:
\vspace{-2mm}
\begin{equation}
    AiA(\mathbf{Q},\mathbf{K},\mathbf{V}) = (\textup{Softmax}\left ( \mathbf{M} + \textup{InnerAttn}(\mathbf{M})) \right )\bar{\mathbf{V}})\mathbf{W_o}
\end{equation}

AiATrack consists of three input branches, the initial frame, the current search frame, and selected intermediate frames. All frames pass through the ResNet50 feature extractor, before performing self-attention in the transformer encoder, which searches for correlation within the same feature array ($Q=K=V$). Thereafter, the feature maps from each branch were combined during the long-term (LT), and short-term (ST) cross-attention modules, thus enable learning across time ($Q\neq K \neq V$). During inferencing, the network incorporated an additional checking algorithm that stores high-performing prior examples (classified by the Dice metric) in its memory, and then call upon them when predicting the current frame by concatenating it to the value. Note that the ResNet50 and the transformer encoders on each branch share their weights.

Similiar to AiATrack, AiAReSeg (in Fig.\ref{fig:AiAReSeg}) consists of three branches, where each branch processes the image through ResNet50. At each feature level of the convolution, the feature channels were connected to the output deconvolution layers via skip connections. The ResNet50 feature maps were processed with intermediate convolutional layers to gradually reduce its channel size, such that it matches the deconvolution layer's intermediate outputs (similiar to UNet++ \cite{zhou2018unet++}). Then, as shown in Fig. \ref{fig:3D deconv}, the feature maps were stacked along a new dimension, generating (H,W,T) sized images, where T represents the time, before it was passed into a 3D convolutional layer to reduce T to 1, prior to further stacking from the next skip connection branch.

In order to adapt AiAReSeg to both aorta and catheter segmentation, we had combined the following loss functions: Dice coefficient (DSC), Binary Cross Entropy (BCE), and Mean Squared Error (MSE), and weighted their importance with factors of 5, 2 and 2, respectively. The Dice loss measures the similarity between predicted and ground truth masks, and encourages models to produce more accurate segmentation results, and handles class imbalance. We also used the BCE loss to assign higher probabilities to the correct class and lower probabilities to the incorrect class, and the MSE to minimise the pixel-to-pixel distance between the ground truth and the prediction.

\section{Experimental evaluation} \label{Experimental evaluation}
We divided the evaluation of this pipeline into two phases: evaluation on hold out simulated image set, and on unseen phantom data. This test examines the capability of the network in generalising to unseen datasets, with the latter being a closer representation of the real patient anatomy.

To evaluate the performance of our system, we selected a handful of common and top-performing detection and segmentation models in literature. Most notably, this includes the Faster-R-CNN and DETR for detection, compared against the performance of AiATrack, while for segmentation we selected the standard UNet, and a clustering based approach, which is explained in Sect. \ref{Section: Benchmarking details}. 

The models were trained and evaluated on both detection and segmentation of the aorta and catheter, evaluated separately. The reason for this choice is that analysis of a larger feature such as the aorta is easier to perform, as it is unique in the input image (with only one duplicate of the cross section if scanned near the aortic arch). Catheters, on the other hand, are significantly more challenging to detect due to the noisy nature of the background, since their shape and intensity range can easily be confused with artifacts or other features, thus affecting performance. In addition, their small size also created significant class imbalance between the feature and the background, making some of the loss metrics highly volatile (such as the Dice loss). 

We evaluated the tracking models using the average precision (AP) metric, defined as the area under the precision-recall curves, evaluated at different intersections over union (IOU) thresholds between 50 and 95$\%$, including their average to form the mean AP score. On the segmentation side, we used the Dice metric (DSC), which indicates the degree of overlap, and the mean absolute error (MAE), which represents the distance from each pixel to its ground truth.
\vspace{-2mm}
\subsection{Training details} 
Experiments were conducted on a workstation with NVIDIA GeForce RTX3080, 32GB RAM, Intel core i7 (10700K). The physics-based simulations were performed on MuJoCo 2.10, where mesh models were decomposed into convex hulls using V-HACD \cite{mammou_vhacd}. The US simulations were generated on the ImFusion Suite (ImFusion GmbH, Munich, Germany), where the ray-casting algorithm \cite{salehi2015patient} was implemented. 8 torso CT images of men and women were selected from the Synapse dataset, and used for catheterisation. Catheterisation simulation was performed for a total duration of 60 seconds, where a data recording of the catheter positions was performed at 4mm intervals along its body to provide a reasonable spatial resolution for reconstruction. During US simulation, the transducer was programmed to follow a predefined spline, and performed a ray-casting simulation at 0.1mm increments along the line. To increase data variability, this spline was also rotated by $\pm 30$  degrees, and re-projected onto the volume surface, creating different viewing angles of the anatomy. Finally, images were divided by sequences into folders, where image/mask pairs that do not contain a catheter were filtered out.

\vspace{-2mm}
\subsection{Phantom data collection details}

A small set of 2D testing images were collected manually in a free-hand manner from a ZONAE Z.One Ultra Ultrasound machine, using a C8-3 (3D) transducer at a scanning depth of 10cm. Axial view scanning was performed on an Elastrat silicon-based aortic arch phantom, immersed in lukewarm saline solution. To mimic a catheterisation procedure, we inserted a Merit Medical 5F vertical catheter at the distal end of the phantom, then followed the tip of the catheter with the US probe. We collected 5 US sequences with varying lengths, ranging between 400 - 700 frames.

\vspace{-2mm}
\subsection{Model specific details}\label{Section: Benchmarking details}

\textbf{AiATrack:} A patch of size $5^2$ was cropped from the frame, and resized into a common dimension of 320 x 320 pixels. A ResNet-50 backbone was used \cite{he2016deep} for feature extraction, downsampling the input until a size of 20x20 was achieved. Each feature map was flattened and passed into the transformer. A 4-head attention module was used, with the inner AiA module reducing the channel dimension of queries and keys to 64. The final prediction head used 3 Conv-BN-ReLU layers, a PrPooling layer and 2 fully connected layers. The model was pretrained for 300 epochs on the LaSOT dataset \cite{fan2019lasot}, then for an additional 200 epochs on the synthetic US dataset, both at a learning rate of $10^{-4}$.

\textbf{DETR:} A standard DETR, with weights pretrained for 500 epochs on the COCO2017 dataset was used in this application \cite{lin2014microsoft}.  The COCO dataset consists of more than 200,000 images consisting of over 80 categories of objects, thus equipping the model with the necessary feature extraction filters. The model is retrained on ultrasound data for 100 epochs. The learning rate in both cases is $10^{-5}$.

\textbf{Faster-R-CNN:}
An implementation of Faster-R-CNN from the Detectron2 library was used \cite{wu2019detectron2}. A ResNet50 backbone was used, together with a feature pyramid network. A COCO2017 pre-trained model was retrained on our own dataset for 100 epochs, with a learning rate of $10^{-5}$.

\textbf{AiAReSeg:}
Our proposed AiAReSeg framework offers training in an end-to-end manner. However, in order to accelerate the process of training, model weights prior to the final segmentation head were initialized with weights from an AiATrack model, pretraiend with 300 epochs on LaSOT at a learning rate of $10^{-5}$.

\textbf{UNet:}
A standard UNet from Ronneberger et al.'s work \cite{ronneberger2015u} was implemented with MONAI \cite{cardoso2022monai} and trained for 100 epochs with a learning rate of $10^{-3}$.

\textbf{Clustering Based evaluation:}
We also designed a partially unsupervised workflow to extract the catheter given a valid aorta segmentation. In this case, an US image was first filtered with the proposed aorta segmentation mask from AiAReSeg, extracting the aorta and its embedded catheter. This patch was then thresholded at a 70$\%$ intensity level, before a K-means clustering was performed (K=2). The final cluster selection was done based on the root mean square variance of each cluster (computed via Eq.\ref{eq:var_rms}, where the cluster with the smallest RMS variance was selected).

\begin{equation} \label{eq:var_rms}
    VAR_{rms} = \sqrt{(var_x)^2+(var_y)^2}
\end{equation}

\begin{table}[t!]
\centering
\vspace{2mm}
\caption{Evaluation on synthetic data: Detection models}

 \resizebox{\columnwidth}{!}{
\begin{tabular}{c|ccc|ccc|}
\cline{2-7}
\multicolumn{1}{l|}{}                       & \multicolumn{3}{c|}{\textbf{Aorta}}                                                                        & \multicolumn{3}{c|}{\textbf{Catheter}}                                                                     \\ \hline
\multicolumn{1}{|c|}{\textbf{Model Name}}   & \multicolumn{1}{c|}{\textbf{AP}} & \multicolumn{1}{c|}{\textbf{AP50}} & \multicolumn{1}{l|}{\textbf{AP75}} & \multicolumn{1}{c|}{\textbf{AP}} & \multicolumn{1}{c|}{\textbf{AP50}} & \multicolumn{1}{l|}{\textbf{AP75}} \\ \hline
\multicolumn{1}{|c|}{\textbf{DETR}}         & \multicolumn{1}{c|}{72.40}       & \multicolumn{1}{c|}{98.80}         & 98.80                              & \multicolumn{1}{c|}{22.56}       & \multicolumn{1}{c|}{ \textbf{77.10}}         & 4.03                               \\ \hline
\multicolumn{1}{|c|}{\textbf{Faster-R-CNN}} & \multicolumn{1}{c|}{89.05}       & \multicolumn{1}{c|}{98.93}         & 98.82                              & \multicolumn{1}{c|}{12.70}       & \multicolumn{1}{c|}{46.60}         & 2.01                               \\ \hline
\multicolumn{1}{|c|}{\textbf{AiATrack}}     & \multicolumn{1}{c|}{\textbf{94.77}}       & \multicolumn{1}{c|}{\textbf{100}}           & \textbf{100}                                & \multicolumn{1}{c|}{\textbf{22.86}}       & \multicolumn{1}{c|}{70.99}         & \textbf{6.33}                               \\ \hline
\end{tabular}}
\vspace{-3mm}
\label{Tab: Results Sim Detection}
\end{table}

\begin{table}[t!]
\centering
\caption{Evaluation on phantom data: Detection models}
\vspace{-3mm}
 \resizebox{\columnwidth}{!}{
\begin{tabular}{c|ccc|ccc|}
\cline{2-7}
\multicolumn{1}{l|}{}                       & \multicolumn{3}{c|}{\textbf{Aorta}}                                                                        & \multicolumn{3}{c|}{\textbf{Catheter}}                                                                     \\ \hline
\multicolumn{1}{|c|}{\textbf{Model Name}}   & \multicolumn{1}{c|}{\textbf{AP}} & \multicolumn{1}{c|}{\textbf{AP50}} & \multicolumn{1}{l|}{\textbf{AP75}} & \multicolumn{1}{c|}{\textbf{AP}} & \multicolumn{1}{c|}{\textbf{AP50}} & \multicolumn{1}{l|}{\textbf{AP75}} \\ \hline
\multicolumn{1}{|c|}{\textbf{DETR}}         & \multicolumn{1}{c|}{1.4}       & \multicolumn{1}{c|}{5.3}         & 0.3                              & \multicolumn{1}{c|}{NA}       & \multicolumn{1}{c|}{NA}         & NA                               \\ \hline
\multicolumn{1}{|c|}{\textbf{Faster-R-CNN}} & \multicolumn{1}{c|}{12.1}       & \multicolumn{1}{c|}{23.7}         & 12.6                              & \multicolumn{1}{c|}{0.9}       & \multicolumn{1}{c|}{5.9}         & 0.1                               \\ \hline
\multicolumn{1}{|c|}{\textbf{AiATrack}}     & \multicolumn{1}{c|}{\textbf{45.7}}       & \multicolumn{1}{c|}{\textbf{100}}           & \textbf{82.62}                                & \multicolumn{1}{c|}{\textbf{14.3}}       & \multicolumn{1}{c|}{\textbf{63.93}}         & \textbf{3.79}                               \\ \hline
\end{tabular}}
\vspace{-3mm}
\label{Tab: Results Phantom Detection}
\end{table}

\begin{table}[t!]
\centering
\vspace{3mm}
\caption{Evaluation on synthetic data: Segmentation models}

 \scalebox{0.9}{
\begin{tabular}{|c|cc|cc|}
\hline
                    & \multicolumn{2}{c|}{\textbf{Aorta}}              & \multicolumn{2}{c|}{\textbf{Catheter}}                        \\ \hline
\textbf{Model Name} & \multicolumn{1}{c|}{\textbf{DSC}} & \textbf{MAE} & \multicolumn{1}{c|}{\textbf{DSC}} & \textbf{MAE}              \\ \hline
\textbf{UNet}       & \multicolumn{1}{c|}{88.95}        & 0.00258      & \multicolumn{1}{c|}{80.06}        & \textbf{0.00010}                   \\ \hline
\textbf{AiAReSeg}   & \multicolumn{1}{c|}{\textbf{91.92}}        & \textbf{0.00213}      & \multicolumn{1}{c|}{\textbf{83.10}}        & 0.00014                   \\ \hline
\textbf{Clustering} & \multicolumn{1}{c|}{-}            & -            & \multicolumn{1}{l|}{46.17}        & \multicolumn{1}{c|}{0.24} \\ \hline
\end{tabular}}
\vspace{-3mm}
\label{Tab: Results Sim Segmentation}
\end{table}

\begin{table}[t!]
\centering
\caption{Evaluation on phantom data: Segmentation models}
\vspace{-3mm}
\scalebox{0.9}{
\begin{tabular}{|c|cc|cc|}
\hline
                    & \multicolumn{2}{c|}{\textbf{Aorta}}              & \multicolumn{2}{c|}{\textbf{Catheter}}           \\ \hline
\textbf{Model Name} & \multicolumn{1}{c|}{\textbf{DSC}} & \textbf{MAE} & \multicolumn{1}{c|}{\textbf{DSC}} & \textbf{MAE} \\ \hline
\textbf{UNet}       & \multicolumn{1}{c|}{32.30}        & 0.037        & \multicolumn{1}{c|}{20.39}        & 0.00068      \\ \hline
\textbf{AiAReSeg}   & \multicolumn{1}{c|}{\textbf{34.11}}        & \textbf{0.032}        & \multicolumn{1}{c|}{\textbf{62.51}}        & \textbf{0.00018}      \\ \hline
\textbf{Clustering} & \multicolumn{1}{c|}{-}            & -            & \multicolumn{1}{l|}{25.80}        & 0.26         \\ \hline
\end{tabular}}
\vspace{-5mm}
\label{Tab: Results Phantom Segmentation}
\end{table}

\vspace{-5mm}
\section{Results} \label{Results}
\vspace{-2mm}
Results from tracking experiments are shown in Tab. \ref{Tab: Results Sim Detection}, which presents the findings for simulations, whereas results from benchtop phantom trials are presented in Tab.\ref{Tab: Results Phantom Detection}. In all cases, AiATrack demonstrated the highest level of mean AP, with a score of 94.77 for aorta and 22.86 for catheter tracking. While AiATrack surpassed all other models across all AP IOU thresholds for aorta tracking, it fell short of the DETR's AP50 of 77.10 (vs 70.99). Nevertheless, it still outperformed the DETR on average. When applied to phantom trials, the DETR and Faster-R-CNN struggled to generalise its performance to these images, with DETR yielding an especially poor mAP performance of 1.4. The same observation was made with catheter detection, where the DETR did not yield any metric for AP, while the Faster-R-CNN's performance was also poor. The AiATrack model's performance far exceeded both cases, at 45.7 and 14.3 for aorta and catheter detection respectively.

Similarly for segmentation, Tab. \ref{Tab: Results Sim Segmentation} presents testing of the model on simulation, and Tab. \ref{Tab: Results Phantom Segmentation} is for phantoms trials. We found that AiAReSeg's performance surpassed UNet in both aorta and catheter segmentation, in simulated (aorta: 91.92 vs 88.95, catheter: 83.10 vs 80.06) and phantom trials (aorta: 34.11 vs 32.30, catheter: 62.51 vs 20.39), indicating that the model was able to generalize to some degree from simulation to reality, without needing to retrain.

% \vspace{-2mm}
\section{Discussions} \label{Discussions}
% \vspace{-2mm}
From these results, we first observe that AiA-based systems yielded the highest level of performance across nearly all detection and segmentation tasks. With simulations, where the texture of generated images were similar to the training data, the detection model performed better on average and at the 50$\%$ and 75$\%$ thresholds. This indicated that within the same image domain, the model surpassed a selection of existing frameworks.  This finding is within our expectations, as the model draws upon temporal information from across the sequences, effectively supplying the knowledge about how this feature is changing over time. 

Furthermore, in neighbouring but different image domains (such as the phantom image domain), although the performance was severely impacted due to lack of retraining, AiATrack still surpassed its competitors, especially in the case of aorta detection, yielding an AP of 100 at 50$\%$ and 82.62 at $75\%$ thresholds. For the more challenging catheter detection task, AiATrack's performance was still higher, in the case that the DETR and the Faster-R-CNN completely failed to generalise. These results indicated the robustness of the AiA framework in adapting to new domains. In the case that the model is provided with a small subset of images from this new domain, it is reasonable to assume that AiATrack will start training with more adapted weights (transferred from previous training examples) to this domain, and require less data to achieve similar levels of performance as Tab. \ref{Tab: Results Sim Detection}.

With segmentation, AiAReSeg used temporal features at the attention and reconstruction level as prior knowledge at different spatial scales to aid it in mask generation. As a result, the AiAReSeg architecture surpassed its UNet competitor in both aorta and catheter segmentation tasks in simulation, and in phantom trials. We recognise that due to the challenging nature of catheter semantic segmentation, where the mask label for each frame typically only consists of 20-100 pixels, the Dice metric is rather harsh in penalising the model, even where the absolute error between the model output and the ground truth is very low. Thus, when we examine the MAE metric, it was also noted that AiAReSeg was significantly better at minimising its distance with the ground truth in the phantom case (0.00018 for AiAReSeg vs 0.00068 for UNet). Considering that catheter localisation in a clinical environment demands high accuracy, we believe that these results demonstrate the potential for our system to perform well when it is sufficiently retrained. 

Finally, the poor performance for phantom aorta segmentation from both models was also investigated, and the main reason found was the significant difference in appearance of the tubular structure in simulation vs in phantom. While our chosen phantom mimics the mechanical properties of an aorta, and its aesthetic appearance, the acoustic behaviour of silicone is very different from reality. As a result, a phantom axial image has high intensity on the top surface of the aorta (indicating high reflectivity), while the lower surface is shadowed, creating a discontinuous tubular shape. This shape was not observed by the model during training using simulated data, hence confusing the networks.

% \vspace{-2mm}
\section{Conclusions} \label{Conclusion} 
% \vspace{-2mm}

In this paper, we presented a solution to the data shortage problem in the field of interventional ultrasound, by presenting a bespoke data synthesis pipeline. Through experimentation, we have demonstrated that the dataset step towards bridging the gap between simulation and reality. Deep learning models trained with this dataset were able to exhibit satisfactory preliminary results on silicon phantoms without needing to retrain. These results pave the way for future works which verifies such models on real patient anatomy. We also present our innovation, the AiAReSeg architecture, which combines temporal information both when attention is applied, and during reconstruction in the 3D deconvolution layers. Injection of temporal information enhanced the model to become a competitive option for catheter segmentation tasks among its rivals. %We hope to improve AiAReSeg in the future, such as by using motion information and other priors to adapt the framework further, leading to a self-supervised or weakly supervised model that can face the challenges of complex medical computer vision tasks.

%                                   % the textheight of the last page by a suitable amount.
%                                   % This command does not take effect until the next page
%                                   % so it should come on the page before the last. Make
%                                   % sure that you do not shorten the textheight too much.

%%%%%%%%%%%%%%%%%%%%%%%%%%%%%%%%%%%%%%%%%%%%%%%%%%%%%%%%%%%%%%%%%%%%%%%%%%%%%%%%

%%%%%%%%%%%%%%%%%%%%%%%%%%%%%%%%%%%%%%%%%%%%%%%%%%%%%%%%%%%%%%%%%%%%%%%%%%%%%%%%

%%%%%%%%%%%%%%%%%%%%%%%%%%%%%%%%%%%%%%%%%%%%%%%%%%%%%%%%%%%%%%%%%%%%%%%%%%%%%%%%

\vfill
\newpage
\bibliographystyle{unsrt}
\bibliography{refs}
\vfill
\newpage

%%%%%%%%%%%%%%%%%%%%%%%%%%%%%%%%%%%%%%%%%%%%%%%%%%%%%%%%%%%%%%%%%%%%%%%%%%%%%%%%

\end{document}